# Who pays? Comparing cost sharing models for a Gold Open Access publication environment[1]

Andre Bruns, Christine Rimmert, Niels Taubert

## Abstract

The article focuses on possible financial effects of the transformation towards Gold Open Access publishing based on article processing charges and studies an aspect that has so far been overlooked: Do possible cost sharing models lead to the same overall expenses or do they result in different financial burdens for the research institutions involved? It takes the current state of Gold OA publishing as a starting point, develops five possible models of attributing costs based on different author roles, number of authors and author-address-combinations. The analysis of the distributional effects of the application of the different models shows that all models result in similar expenditures for the overwhelming majority of institutions. Still, there are some research institutions where the difference between most and least expensive model results in a considerable amount of money. Given that the model calculation only considers publications that are Open Access and where all authors come from Germany, it is likely that different cost sharing models will become an issue in the debate on how to shoulder a possible large scale transformation towards Open Access based on publication fees.

## Introduction

In a publication environment that is based on the subscription model, the question 'who pays?' can be answered easily: In general, it is the library with its duty to provide access to publications for a particular group of users. Libraries can choose between individual subscriptions, bundle deals or participate in consortia, but it is clear that, at the end of the day, they have to pay for the subscription they signed. In the course of the transformation towards Gold Open Access, which today is a top priority for a number of influential players in science policy[2], this situation changes as the costs for publishing have to be covered by other sources. Journals may be founded and run on the initiative of the scientific community or may be sponsored by funding agencies, academies or learned societies. Journals that do not charge any money from anyone are sometimes called platinum (Wilson 2007) or diamond OA (Fuchs & Sandoval 2013). Nevertheless, business models that are most prominently discussed in the context of an OA transformation do apply publication fees, most often called article processing charges (APC). APC are being used for full OA journals that provide free access to all publications, in the context of so-called hybrid journals which provide the option to pay a fee to make a particular article freely accessible within an otherwise subscription-based journal, and also for publish and read agreements like, for example, the DEAL contracts (Kupferschmidt 2019). There is a strong consensus that not the author but his or her institution should pay for the publication (Suber

---

[1] This work was supported by the Deutsche Forschungsgemeinschaft (DFG) [grant number WI 3919/1-1].

[2] Besides other activities one should mention the block grants of Wellcome Trust and the RCUK for Open Access as a response to the Finch report (Finch 2012), the open access policy of the European Union in the context of FP/ and Horizon 2020 and the publish and read contracts in the context of DEAL.

2012: 138, Björk/Solomon 2014: 13). Regarding the transformation towards Gold OA, some issues have already been studied.

One question a number of studies address is to what extent the publication output of an entity of a research system (e.g. institutions, countries, disciplines) is freely available online in the formal communication channel (Laakso et al. 2011; Gargouri et al. 2012; Archambault et al 2013; Crawford 2015; Wohlgemuth et al. 2017; Piwowar et al, 2018, Abendiyarandi and Mayr 2019). Studies differ with regard to the databases, the sources of OA evidence, and the definition of OA types (Martín-Martín et al. 2018). Nevertheless, there is some evidence that can be found across all contributions: first, the share of publications that are freely available online in the formal communication channel has reached a level that can hardly be overlooked and that today contributes to the supply of information within many disciplines and fields within science. In addition, the dynamics of growth of the gold OA share still sustains and is today mainly driven by science policy on the national as well as on the international level.

Given that Open Access provided by journals is developing very dynamically, several studies aim to understand the economics of the publishing model. Many OA journals are based on subsidies, but a large number of journals, in particular those of the commercial publishers, apply APC. Early prominent examples are *BioMedCentral* and the *Public Library of Science* which adopted publication fees already around 2000 (Solomon and Björk 2012; Björk and Solomon 2015). A first set of studies report to what extent the business models of OA-journals rely on APC. On the global level, Morrison et al. (2015) find that more than two thirds of the journals included in the Directory of Open Access Journals (DOAJ) apply publication fees. The application of APC seems to differ by field (Crawford 2017). Medicine is, for example, a field with low adoption as two thirds of the journals refrain to impose APC (Asai 2019). In addition, the take-up of APC also varies by region. Latin America, the Middle East, and Eastern Europe have a large share of OA journals that do not charge APC (Crawford 2017) and are financed by other means, such as subsidies from the state as in the case of Brazil (Appel and Albagli 2019).

Another set of studies is interested in the price for publishing in an APC environment. Because of the lack of other data, early studies referred to list prices on publishers' websites (Morrison et al. 2015) or to prices as recorded by DOAJ (Björk and Salomon 2015). Given that the amount of money that is actually paid for APC can differ from list prices, and given that even payments for articles that are published in the same journal may also vary, more recent publications are based on collections of actual payments like, for example, the OpenAPC initiative (Pieper and Broschinski 2018). Five results of the APC prices/payment studies seem to be worth highlighting. First, the average price/payment reported in the studies vary at similar scale: Björk and Salomon (2015) report a mean of 1,255 USD, Morrison et al. (2015) 1,221 USD as average list price by journal and Asai (2019) 1,018 USD average APC list price for medicine. Studies based on recorded payments show similar results: Jahn and Tulley (2016) calculate 1,298 € average payments for APC in full open access journals and Pieper and Broschinski (2018) an average of 1,479 €  Second, all studies report large standard divisions indicating that there is much variance of the pricing of APC. Third, some studies observe an increase of average price/payments (Jahn and Tullney 2016; Pieper and Broschinski 2018). Fourth, there is some evidence that APC are higher for hybrid journals than for full open access journals (Solomon and Björk 2012; Pinfield et al 2015; Jahn and Tulley 2016, University of California Libraries 2016). Fifth, the amount of APC seems to vary by disciplines (Solomon and Björk 2016).

The large variations raise the question as to what factors determine the pricing of APC. Björk and Solomon (2015) found a moderate relation between price and quality, taking the Source normalized Impact per paper (SNIP) as a proxy for quality. Asai (2019) reported the type of publisher, the language of the journal and the average number of citations as determinants, while Schönfelder (2020) found that quality (as measured by SNIP) and whether a journal is a full OA or hybrid are the most important determinants.

A last tier of studies that should be mentioned here ask about the financial effects of an ongoing open access transformation on the level of institutions. Besides rough and not very reliable estimations on a global scale (Schimmer et al. 2015) they analyze possible financial risks of an OA transformation for the institutions involved. An important contribution is the "Pay it forward"-study that focuses on research-intensive institutions in North America. It starts with the hypothesis that a transformation towards an APC-based model would relieve libraries budget of universities with a rather low publication output and would put pressure on libraries of research-intensive universities regarding their budget. A so-called APC break-even point was calculated for 13 North American universities. In theory, this is the amount of money that would be available for each publication of a given university if all expenditures for subscriptions would be spent on APC. The APC break-even point is as low as 533 USD for UC San Francisco, 709 USD for Harvard and 843 USD for UC Los Angeles (University of California Libraries 2016). A study calculating the break-even-point for German universities came to the conclusion that, taking the current level of APC as reported by OpenAPC as a threshold value, four of them would have enough money to cover APC for all publications, while one university is close to the current level of average payments for APC (Taubert 2019).

Besides these studies, there is not a lot of robust information about other possible financial effects of a large scale transformation towards APC-financed OA. This paper therefore focuses on one of the aspects that have – up to now – not attracted much attention, namely, the question as to how costs for APC are being attributed to institutions involved in a publication. Usually it is assumed that APC should be covered by the institution of the reprint author. This holds for APC paid by central publication funds as well as in the context of publish and read contracts. However, there is no natural way of attributing costs, especially in the case of co-authored articles with authors from different research institutions. The question 'who pays?' can be answered differently as costs can be shared in many ways.

This article aims to shed light on different possibilities of sharing costs for the publication in APC-based journals and compares the financial effects between them on the level of institutions. Do possible cost sharing models lead to the same overall expenses or do they result in different financial burdens for research institutions? To answer this question, a model calculation is undertaken. It takes the current state of Gold OA publishing as a starting point, develops five possible models of attributing costs and asks about possible financial effects of their application. Even though the analysis only focuses on the German research system for practical reasons, the methods and results are also of relevance for other countries.

The paper is organized as follows: In a first step the data and methods of the analysis are outlined. This is followed by an examination of what kind of document types APC are actually paid for. In the third step the overall publication output in Gold OA journals of different types of institutions is described. The fourth step provides an analysis of the structure of co-authorships in different sectors of the German research system in order to determine to what extent of the publication output the cost sharing models can be applied. Possible cost sharing models are introduced in a fifth step, followed by the analysis of the distributional effects of their application. The article concludes with some considerations about the application of cost sharing models in a Gold OA publishing environment.

## 1. Data and Methods

A model calculation that investigates possible distribution effects of the application of different cost sharing models requires the combination of a publication database that allows to identify the publication output on the level of research institutions, a data source that determines the OA status of a publication, and a source that provides cost information for APC that are actually paid.

*Publication database*

First, publication data covering a well-defined section of the publication output of German research institutions are taken from the Web of Science (WoS) bibliometric database provided by the Competence Centre for Bibliometrics[3] in its version of May 2018.[4] The main reason for choosing this version of the database is that it is enriched with disambiguated institutional addresses (Rimmert et al. 2017), allowing a precise identification of the publication output of a research institution. Given that disambiguated address information are available for Germany only, the model calculation is restricted to that country.

*Gold OA evidence source*

Second, the analysis requires confining the share of the publication output of German research institutions in Gold OA journals. According to Archambault et al. (2014), Gold OA means that a journal provides immediate free online access to all of its publications (cover-to-cover). The WoS now provides information about the Gold OA status based on the Directory of Open Access journals (DOAJ).[5] This source turned out not to be exhaustive for the WoS[6]. Therefore, we use the ISSN-Gold-OA 3.0 list (Bruns et al. 2019) which merges DOAJ, PubMed Central (PMC) journal lists[7] and the Directory of Open Access Scholarly Resources (ROAD)[8] for the identification of Gold OA journals.

---

[3]  http://www.forschungsinfo.de/Bibliometrie/ , November 07[th] 2019.

[4]  Access to the raw data is provided by the Competence Centre for Bibliometrics
 (http://www.bibliometrie.info, November 07[th] 2019.)

*APC cost information*

Third, APC that are actually paid for publications in Gold OA journals differ in many cases from the list prices indicated by the publishers (Solomon/Björk 2016: 5).[9] Therefore, costs for APC were actually paid by institutions that were identified by using the OpenAPC (OAPC) dataset of the INTACT project.[10] To our best knowledge, OAPC is the largest collection of fees actually paid for articles in Gold OA as well as in hybrid journals.[11] It includes APC payments from research institutions in Austria, Canada, Czech Republic, Finland, Germany, Great Britain, Greece, Hungary, Italy, Norway, Qatar, Spain, Sweden, Switzerland, and the United States.[12]

## 2. Document types article processing charges are paid for

The term 'article processing charges' suggests that such a fee has to be paid for 'research articles' only, but this assumption might not be true. In the first step it is analyzed whether APC are actually paid for any other document types besides 'articles' to determine what publication types should be included in the analysis. Therefore, the OAPC dataset is matched with publication data taken from the WoS to add the document type. 71,610 of 78,263 publications of the OAPC dataset can be identified in the WoS.

Table 1 should be read as follows: The column 'in % of OAPC' shows the share of the respective document type in the OAPC database, while the column 'in % of WoS' reports the share of the document type in the WoS. As expected, the overwhelming majority in OPAC are 'articles' but there are also other document types for which APC are paid. First, there is a considerable share of 'reviews', the only document type with a share larger than 1% of the OAPC database. Second, the occurrence of small shares of 'meeting' and 'proceedings paper' needs some considerations. These documents are additionally assigned to the document type 'article' in the WoS, and the share of these document types are much larger in the WoS than in the OAPC dataset. This finding suggests that there is only a small fraction of 'meeting' and 'proceedings paper' that apply an APC-business-model. Third, the analysis shows that the APC model is also applied for small fractions of 'editorials', 'editorial materials' and 'letters'. Given that their overall frequency is low in the WoS, the number of cases where APC are paid is also small. All other document types play almost no role. One should note that shares do not sum up to 100% as publications can be assigned to more than one document type.

---

[5] https://doaj.org/, November 07th 2019.

[6] Compared with the ISSN-Gold-OA list, DOAJ identifies only 64% ISSNs of Gold OA journals in the WoS database (Wohlgemuth et al. 2016).

[7] https://www.ncbi.nlm.nih.gov/pmc/journals/, November 07th 2019.

[8] https://www.issn.org/the-issn-international-is-pleased-to-introduce-road/, November 07th 2019. ROAD data can be accessed after registration.

[9] For a study based on APC-information taken from the public internet, see Morrison et al. (2015).

[10] For a more detailed description of the dataset, see Pieper/Broschinski (2018) and https://www.intact-project.org/openapc/, November 07th 2019.

[11] Subscription-based journals allow authors to make their individual article immediately available online after APC have been paid (Prosser 2003).

[12] On July 4th 2019 it contained cost information for 78,263 hybrid open access and gold open access publications.

| document type (WoS) | # publ. in OAPC | in % of OAPC | in % of WoS | included |
|---|---|---|---|---|
| Article | 66,574 | 92.97 | 74.42 | Y |
| Review | 4,214 | 5.88 | 5.80 | Y |
| Meeting | 522 | 0.73 | 7.88 | N |
| Proceedings Paper | 522 | 0.73 | 7.85 | N |
| Editorial Material | 477 | 0.67 | 3.67 | Y |
| Editorial | 477 | 0.67 | 3.67 | Y |
| Letter | 286 | 0.4 | 1.87 | Y |
| Data Paper | 46 | 0.06 | 0.03 | Y |
| Correction | 29 | 0.04 | 0.81 | Y |
| Unspecified | 15 | 0.02 | 0.01 | Y |
| Book | 13 | 0.02 | 0 | N |
| Book Chapter | 13 | 0.02 | 0 | N |
| Abstract | 11 | 0.02 | 6.35 | N |
| Meeting Abstract | 11 | 0.02 | 6.35 | N |
| Book Review | 4 | 0.01 | 1.31 | Y |
| Database Review | 3 | 0 | 0 | Y |
| Software Review | 3 | 0 | 0.01 | Y |
| News | 2 | 0 | 0.49 | Y |
| Retracted Publication | 2 | 0 | 0.01 | Y |
| News Item | 2 | 0 | 0.49 | Y |
| Retraction | 1 | 0 | 0.02 | Y |

*Table 1: Document types in OAPC*

# 3. Publication output of German research institutions in Gold OA journals

In the next step we will provide an overview of the publication output of German research institutions in Gold OA journals. The focus is on publications covered by the WoS and includes all documents indicated in column 'included' in table 1 with at least one author from a German research institution. For the identification of these publications disambiguated institutional addresses were used (Rimmert et al. 2017). One remarkable feature of the publicly funded German research system is that it consists of five main pillars or sectors: 'Universities' (Univ.), 'Helmholtz Association' (HGF), 'Fraunhofer Gesellschaft' (FhG), 'Leibniz Association' (WGL) and 'Max-Planck-Society' (MPG) (Heinze/Kuhlmann 2008: 890ff.). In some cases publications cannot be assigned to individual research institutions but to the level of sectors only. Given that the analysis refers to the level of research institutions, those assignments were excluded. The ISSN-GOLD-OA list 3.0 was used for the identification of publications in Gold OA journals and the analysis was restricted to publication years between 2014 and 2018.

| Publication Year | # Gold OA publications | % of Gold OA publications |
|---|---|---|
| 2014 | 14,718 | 10.90 |
| 2015 | 16,593 | 11.91 |
| 2016 | 19,559 | 13.36 |
| 2017 | 21,334 | 14.76 |
| 2018 | 22,097 | 16.16 |
| Sum | 94,301 | |

*Table 2: Gold OA publications with authors from Germany in WoS, by year*

Table 2 provides an overview of the results. 94,301 publications with addresses from 361 German research institutions were published in Gold OA journals covered by WoS. There is a remarkable increase in the total number of publications in the period analyzed here. Table 3 breaks down the results into sectors. As research institutions may belong to more than one sector[13] and publications can be jointly published by authors with affiliations of institutions from different sectors, the sum of the publications for all sectors are larger than the total number of publications reported in table 2. The large differences of the average number of publications in Gold OA journals are predominantly a result of a differing size of the research institutions within the sectors and – in particular in the case of FhG – of different missions of the institutions within the German research system.

---

[13] Examples are the Max Planck Institute for Plasma Physics which also belongs to the MPG, HGF, and the University of Greifswald and KIT Karlsruhe which are both related to Universities and HGF.

| Sector | # research institutions | Average publ. count (Gold OA) per institution (arithmetic mean) |
|---|---|---|
| Universities | 107 | 2821.72 |
| MPG | 85 | 266.84 |
| WGL | 85 | 190.25 |
| FhG | 65 | 64.43 |
| HGF | 21 | 1507.15 |

*Table 3: Gold OA publications with authors from Germany in WoS, by sector*

The distribution of the publications in gold OA journals with addresses from German research institutions over different document types are reported in table 4. 'Article' is by far the most frequent document type with 88%, followed by 'review' (6%). The document types 'editorial' and 'editorial material' are distributed evenly (2.6%), and the same goes for 'meeting' and 'proceedings paper' (0.9%). All other document types are far below 1% of the publication output.

| Document type | # Gold OA publications | in % |
|---|---|---|
| Article | 84,439 | 88.95 |
| Review | 5,659 | 5.96 |
| Editorial Material | 2,482 | 2.61 |
| Editorial | 2,482 | 2.61 |
| Correction | 836 | 0.88 |
| Letter | 832 | 0.88 |
| Meeting | 546 | 0.58 |
| Proceedings Paper | 546 | 0.58 |
| Data Paper | 177 | 0.19 |
| Book Review | 174 | 0.18 |
| Unspecified | 47 | 0.05 |
| Retraction | 6 | 0.01 |
| Retracted Publication | 3 | 0 |
| Book Chapter | 3 | 0 |
| Book | 3 | 0 |
| Film Review | 2 | 0 |
| Hardware Review | 1 | 0 |

*Table 4: Gold OA publications with authors from Germany in WoS, by document type*

## 4. Co-operations of German research institutes

In the next step we analyze co-operations as indicated by the structure of co-authorship. The goal is to identify the fraction of publications where the model calculation can be practically applied and where the application of different cost sharing models may also result in different financial contributions for the institutions involved. Therefore, five different types of authorship are distinguished, and each publication can be attributed to (exactly) one of them.

1. *$K_0$*: Publications from only one German research institution (either single authored or inner institutional co-operation) without any participation of foreign institutions.
2. *$K_NSec$*: Publications with authors from at least two or more German institutions from the selected sectors[14] but without any participation of foreign institutions.
3. *$K_N$*: Publications without any participation of foreign institutions but with authors from two or more German institutions with exactly one research institution from the selected sectors (without *$K_NSec$*).
4. *$K_I$*: Publications with one or more author(s) from non-German institutions and with German authors from only one institution of the selected sectors (without *$K_ISec$*).
5. *$K_ISec$*: Publications with one or more authors from foreign institutions and German author(s) from at least two different institutions of the selected sectors.

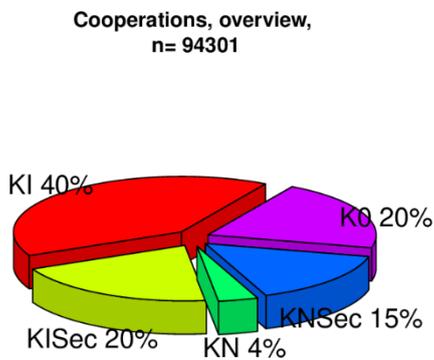
*Figure 1: All sectors*

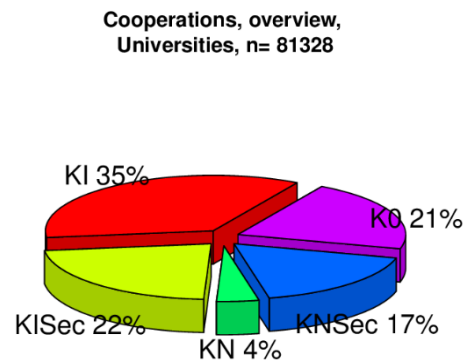
*Figure 2: Universities (Univ.)*

---

[14] Universities, MPG, HGF, WGL, FhG

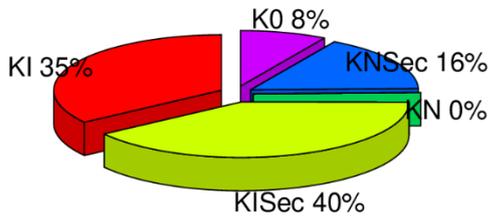

*Figure 3: Max Planck Society (MPG)*

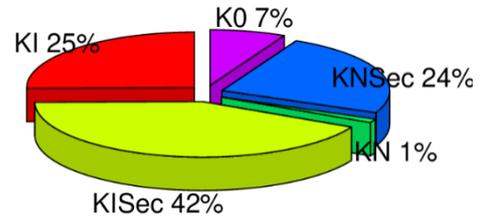

*Figure 4: Helmholtz Association of German Research Centres (HGF)*

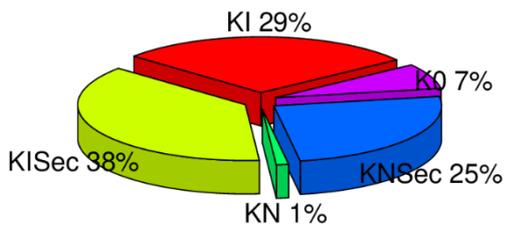

*Figure 5: Leibniz Association (WGL)*

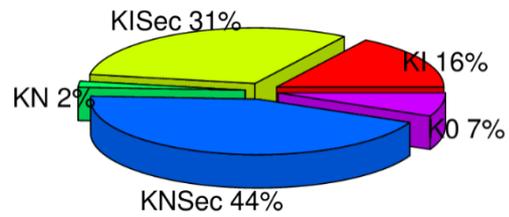

*Figure 6: Fraunhofer Society (FhG)*

Figures 1-6 show the distributions of the five cooperation types for the entire German research system and for all sectors.[15] Given that the major share of German publications involves authors from a university, the overall distribution is largely influenced by the distribution in the university sector. Three additional results are eye-catching: first, universities are the only type of institution with a notable share of single authored publications or in-house cooperation ($K_0$: 21%). Second, the lion's share of publications (varying between 67-75%) from MPG, HGF and WGL are a product of international co-operations (sum of $K_I$ and $K_I Sec$). Third, a large share of publications from the Fraunhofer Gesellschaft results from national cross-sectoral co-operations. In other words: Authors from Fraunhofer institutes tend to collaborate with research institutes of other sectors.[16]

## 5. Cost sharing models

When it comes to the question how costs for article processing charges can be shared between the institutions involved in a publication, one can imagine many possible answers. In addition, there are multiple scenarios on how the publication market might develop and how the APC model will be adopted in different settings. In the case of models that require cost sharing between different institutions, one can, for example, imagine a situation where all institutions show solidarity and contribute to the costs, as well as situations where a smaller or larger number of institutions decide not to participate and free ride. To reduce the complexity, five basic cost sharing models are introduced without any further assumption about possible scenarios in which they are applied. The models represent different criteria of relevance as well as different ideas of fairness regarding the coverage of the costs of publications.

1. *First author model:* In this model, the institution of the first author covers the entire costs for APC. Such a model can be justified with reference to the mechanisms of the attribution of merits within science: An achievement is usually attributed to the first author.[17] An advantage of the first author model is that it already would work in part, even if only a few research institutions would apply it.
2. *Reprint author model:* In this model, the institution of the reprint author covers the entire costs of a publication. This model is favored by OA protagonists (Schimmer et al. 2015: 7) as well as publishers.[18] One peculiarity of the model is that there are cases of publications with more than one reprint author. In these cases, the model would work if the authors manage the attribution of costs. For the model calculation that is undertaken here, the costs will be split up and attributed to all institutions with a reprint author. Again, an advantage of the model is that it would work in part, even if only some research institutions would decide to apply it.

---

[15] The sum of the co-operations attributed to different sectors is larger than the number of overall co-operations because inter-sectoral co-operations that count for all sectors involved.

[16] This result reflects the mission of the Fraunhofer Gesellschaft in the German research system to conduct "innovation-oriented research for the benefit of private and public enterprise" (Cuhls et al. 2012: 233).

[17] The attribution is similar to the mechanisms of attributing merits suggested by Cole/Cole (1973).

[18] See e.g. Springer (https://www.springeropen.com/get-published/article-processing-charges) and Wiley (https://authorservices.wiley.com/open-research/open-access/for-authors/faqs.html), November 07th 2019.

3. *Institutions contribute equally*: In this model, costs for APC are equally shared between all institutions mentioned as affiliations of authors. 'Equal' means that all institutions contribute with the same amount. The rationale is that other authors than the first or the reprint author also benefit from a publication and that their institutions therefore contribute to the costs. A disadvantage of the model is that it only works if all institutions involved in a publication agree to contribute. If some of them decide to free ride, the financial burden would be higher for the remaining institutions.

4. *Institutions contribute, weighted by the number of authors:* In this model, all institutions mentioned as affiliations contribute to the cost of a publication, but the individual amount depends on the number of authors from the particular institution. In other words, the contribution of an institution is calculated on the ground of (fractional) authorships (van Hooydonk 1997: Aksnes et al. 2012: 37; Perianes-Rodriguez et al. 2016). The model takes into account that in larger co-operations many authors come from the same institution while other institutions might be represented by a single or few authors only. Again, a disadvantage of the model is that it only works if all institutions involved agree on it.[19] This cost sharing model can be calculated in two different ways. The share of an institution can be weighted by author-institution-combination (4a) or by authors (4b).

After this brief introduction a fictitious example and formal definition of the different cost sharing models are given. Assume that publication p has the following authors and affiliations:

| Author | Institution |
|---|---|
| $A_1$ | $I_1$ |
| $A_2$ | $I_2$ |
| $A_2$ | $I_3$ |
| $A_3$ | $I_4$ |
| $A_4$ | $I_4$ |
| $A_5$ | $I_4$ |

*Table 5: Authors and Institutions of a fictitious publication*

It is also assumed that $A_1$ is first author and reprint author. The counting methods of the different cost sharing models are summarized in table 6.

---

[19] The cost sharing model based on the fractional authorship and taking authors as a basic unit is similar to the fractional counting of the publication output in bibliometric analysis (see Perianes-Rodriguez et al. 2016).

| No | Description | For institution $I_x$ in publication $p$ $share_p(I_x)$ |
|----|-------------|---------------------------------------------------------|
| 1 | APC fully paid by first author's institution(s) | if $I_x$ = first author inst.: $\frac{1}{\text{\# first author institution (s)}}$ <br> else 0 |
| 2 | APC fully paid by reprint author's institution(s) | if $I_x$ = RP author inst.: $\frac{1}{\text{\# reprint author institution (s)}}$ <br> else 0 |
| 3 | equal shares for institutions | $\frac{1}{\text{\# institution(s)}}$ |
| 4a | shares for institutions weighted by authors, basic unit: author-institution-combination | $\frac{\text{\# author} - \text{institution} - \text{combination in } I_x}{\text{\# author} - \text{institution} - \text{combination}}$ |
| 4b | shares for institutions weighted by authors, basic unit: author | $\sum a \in \text{authors of } I_x \text{ in } p \; \frac{1}{\text{\# inst for a}} * \frac{1}{\text{\# authors}}$ |

*Table 6: Counting methods applied for sharing models*

Table 7 gives an overview of the results of the different cost sharing models when applied to the fictitious publication. It is evident that they result in different costs for each of the four institutions.

| Inst. | first author (1) | reprint author (2) | equal shares (3) | author weighted, basic unit author-inst.-comb (4a) | author weighted, basic unit authors (4b) |
|-------|------------------|---------------------|------------------|----------------------------------------------------|------------------------------------------|
| $I_1$ | 1 | 1 | $\frac{1}{4}$ | $\frac{1}{6}$ | $\frac{1}{5}$ |
| $I_2$ | 0 | 0 | $\frac{1}{4}$ | $\frac{1}{6}$ | $\frac{1}{10}$ |
| $I_3$ | 0 | 0 | $\frac{1}{4}$ | $\frac{1}{6}$ | $\frac{1}{10}$ |
| $I_4$ | 0 | 0 | $\frac{1}{4}$ | $\frac{3}{6}$ | $\frac{3}{5}$ |
| Sum   | 1 | 1 | 1 | 1 | 1 |

*Table 7: Comparison of the results of different cost sharing models (example publication)*

## 5.1 First author model

In the following some background information is provided for each model to deepen the understanding of the logic of attributing costs. To begin with the cost sharing model based on the first authors' affiliation, it would only result in similar financial burdens if all institutions have a similar first author share (number of publications with first author divided by the total number of publications). Figure 7 shows that the first author share differs among German research institutions. The cost sharing model would disadvantage those institutions with a first author share higher than the average 66.9%. Given that the number of OA publications is low for some institutions resulting in outliers, the analysis was restricted to institutions with an output > 50 publications in OA journals between 2014 and 2018. The lower graph in figure 7 points to a more narrow distribution for these universities with an average first author share of 68.3%.

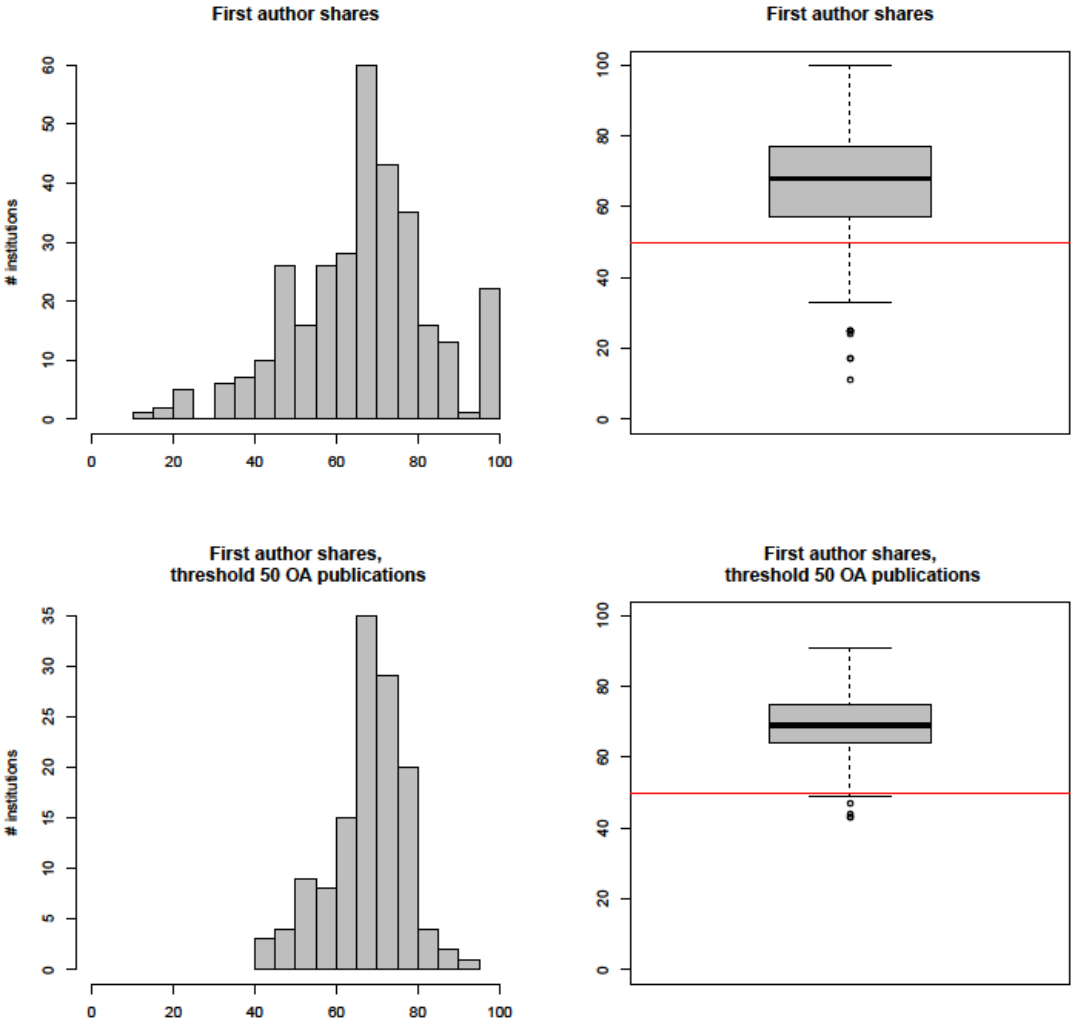

*Figure 7: First author shares for institutions*

## 5.2 Reprint author model

A model where costs are attributed on the ground of the reprint authors' affiliation would only result in similar financial burdens if all institutions have a similar share of reprint authors (number of publications with reprint author divided by the total number of publications of the institution). Figure 8 shows that the reprint author shares differ among institutions and that the model would disadvantage those institutions with a reprint author share higher than the average 61.5%. Given that the number of OA publications is low for some institutions in the respected publication period resulting in outliers, the analysis was again restricted to institutions with more than 50 OA publications between 2014 and 2018. The lower graphs in figure 8 show a more narrow distribution with an average reprint author share of 62.7%.

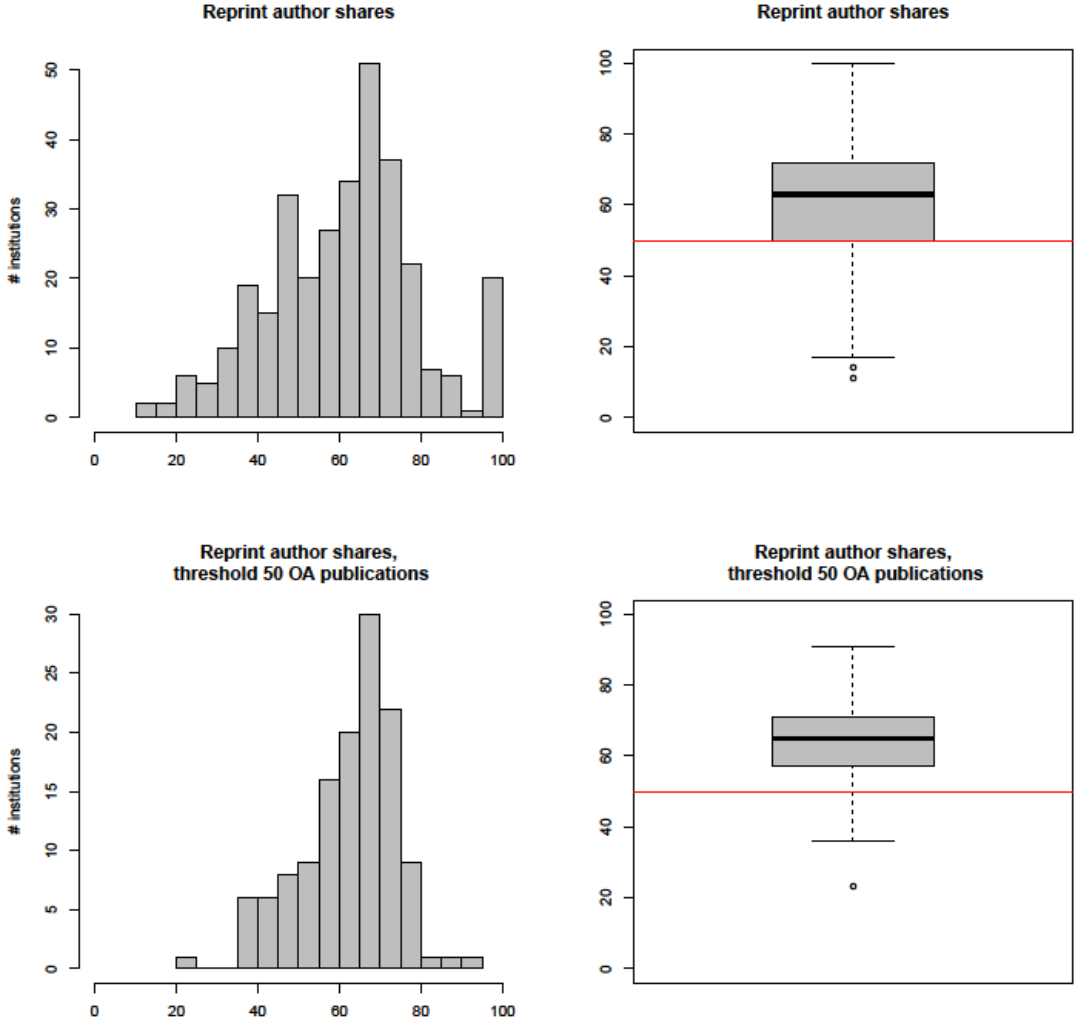

*Figure 8: Reprint author shares for institutions*

## 5.3 Institutions contribute equally

In this model, costs for APC are shared between institutions, and the financial burden for an institution depends on the average number of additional affiliations of co-authors. Given that disambiguated address information is available for German institutions only, the statistical description of the model is restricted to publications without international participation. The distribution shows that 15.3% of these publications come from one institution (KN which can be treated like K0 as the definition implies that there is only one institution of the selected sectors in KN which pays for the article) and in these cases the distribution of costs are similar to the distribution of the first two models. In 84.7% of these publications the 'institutions contribute equally'-model will result in a different attribution of costs than the first two models.

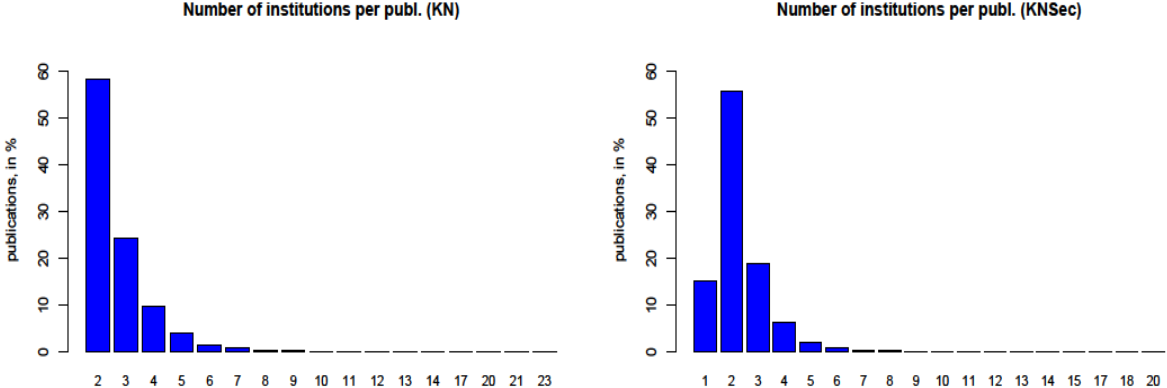

*Figure 9: Number of collaborating German institutions, cross-sectoral and inner-sectoral cooperations*

## 5.4 Institutions contribute, weighted by the number of authors

In this model, the share of costs for a particular institution would depend on the number of institutions involved in a publication weighted by the number of authors that belong to the same institution. Figure 10 gives an overview of the number of authors per publication. Figure 11 shows the distribution of the number of authors per publication for each of the five main sectors of the German research system.

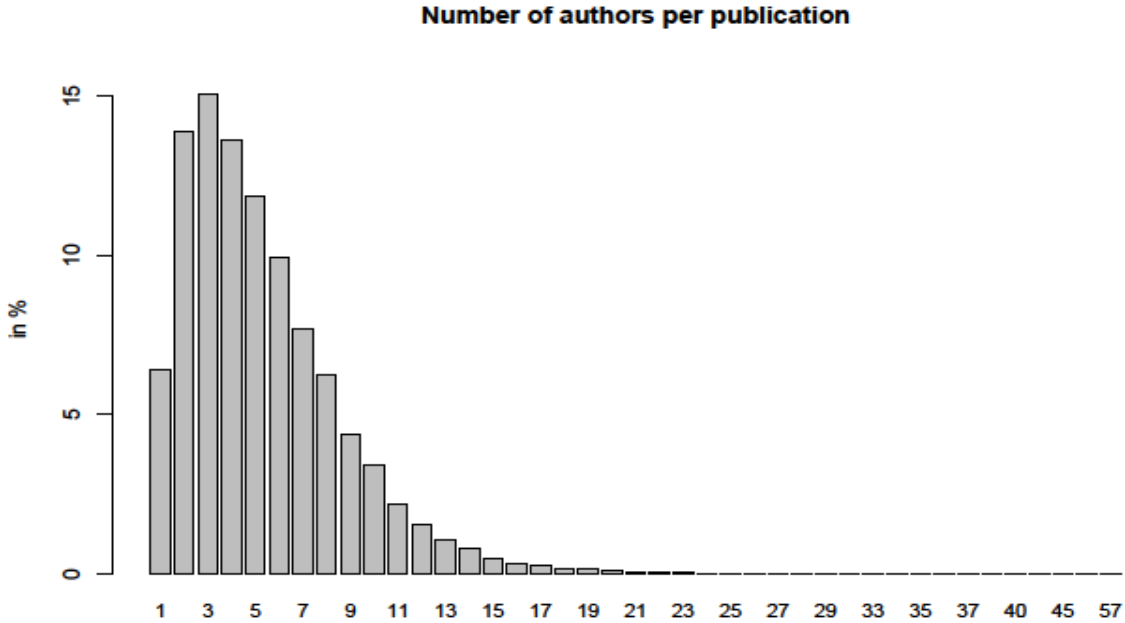

*Figure 10: Number of authors per publication*

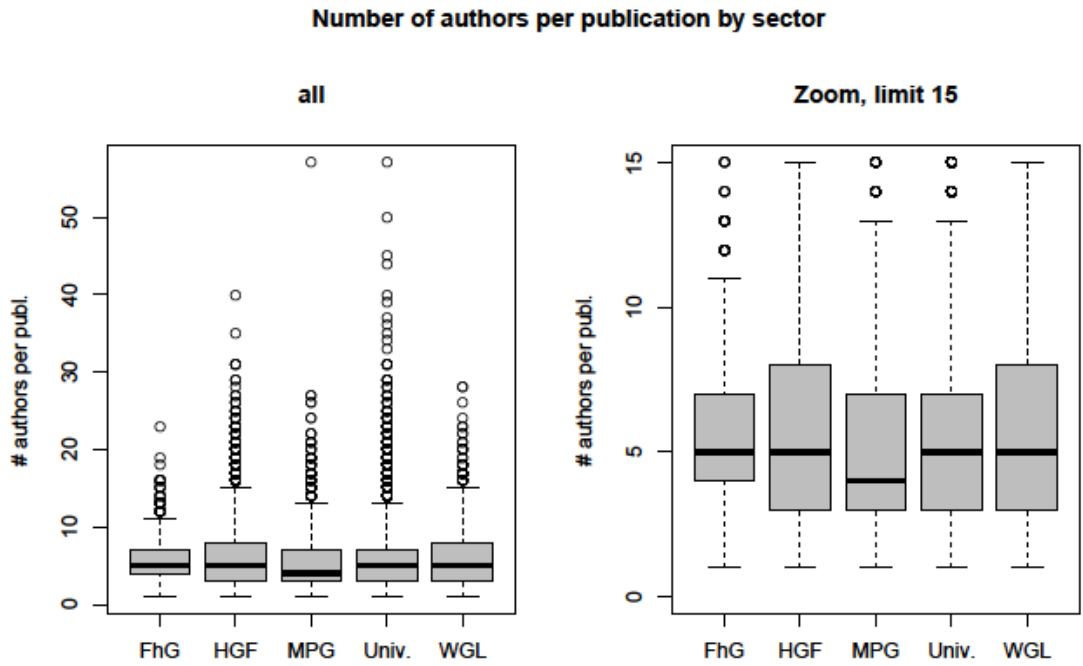

*Figure 11: Number of authors per publication (selected sectors)*

By far, the majority of the publications are co-authored. Only 6.4% of the publications have a single author. In most cases the size of the author team tends to be small. 70.8% of the publications have less than seven authors.[20] In addition, the box plot limited to 15 authors shows that the Helmholtz Association, as well as the Leibniz Association, tends to have larger author teams than the other research institutions. This result reflects the mission of the Helmholtz Association, which runs large scale research infrastructures and conducts 'big science' (Hohn 2010: 460; 2016: 554).

## 6. Comparison of cost sharing models

After having introduced the different cost sharing models, we now compare their financial effects on the level of institutions. This step aims to answer the questions if and to what extent their application results in different financial burdens. The analysis is limited with regard to four aspects: first, and as already described, it includes publications that are covered by the WoS only. Second, it includes single authored and co-authored publications from a single German institution ($K_0$) as well as co-authored publications where all authors are affiliated to German institutions ($K_N$, $K_N$Sec). Publications of international co-operations ($K_I$, $K_I$Sec) are not included. This restriction had to be made as the disambiguated address information is available for Germany only. Therefore, the number of institutions involved in international institutions cannot be determined on that ground.[21] Third, the analysis was restricted to publications where the first author and the reprint author are affiliated to one of the selected sectors. Fourth, the analysis is restricted to publications in Gold OA journals. The rationale is to compare how actual costs can be distributed and not how the models apply in a hardly foreseeable future.

---

[20] Very large author teams consisting of more than 1,000 authors are a unique feature of publications in 'physics' (Huang 2015: 2138). It becomes apparent from figure 11 in which research sectors members of these groups are located.

The results of the comparison are given in 'publication units'. A 'publication unit' (PU) is defined as a full publication an institution has to pay for. Multiplied with the average APC taken from the OAPC dataset, the overall cost for all publications of an institution can roughly be estimated. For the year 2018, the average APC for publications in Gold OA journals paid by German research institutions is 1,540€[22]. Figure 12 provides an overview of the comparison. For each institution (x-axis) the publication units are given for each cost sharing model sorted by the total number of open access publications. Institutions that do not belong to one of the sectors of the German research system were excluded from the analysis.

---

[21] For that reason, fractional cost sharing models (3, 4a, and 4b) cannot be calculated for international publications.

[22] See: https://treemaps.intact-project.org/apcdata/openapc/#institution/period=2018&is_hybrid=FALSE&country=DEU, November 07th 2019.

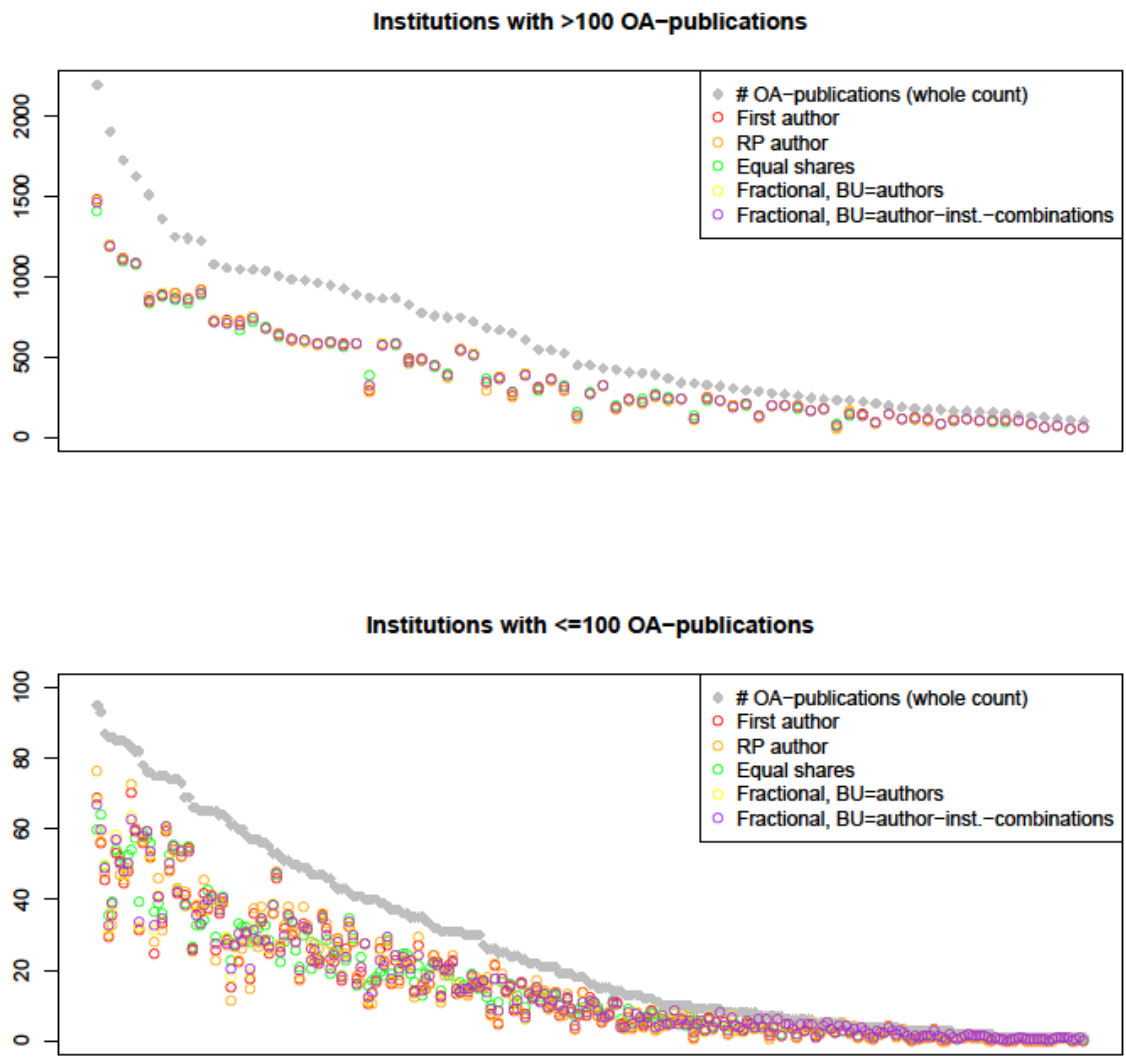

*Figure 12: Values per institution for different counting methods, split in institutions with less equal and more than 100 OA-publications*

Figure 12 indicates that the cost sharing models would result in different financial burdens for each German research institution, although the differences do not seem to be very large in most cases. To investigate the effects in greater detail, the differences between the model resulting in minimum and maximum publication units are calculated for each university. The box-plots in figure 13 show the results. The median (2.48 publication units) and small stretch of the box (including 50% of the institutions) confirm the initial impression that there are small financial effects for most of the research institutions between the cost sharing model with the lowest and the highest costs. For the five year period 2014-2018, a difference of 2.48 publication units in Gold OA journals would roughly mean a difference of 3,819€ between most and least favorable cost sharing models. Nevertheless, there are a number of outliers with a larger difference between the two most diverging models.

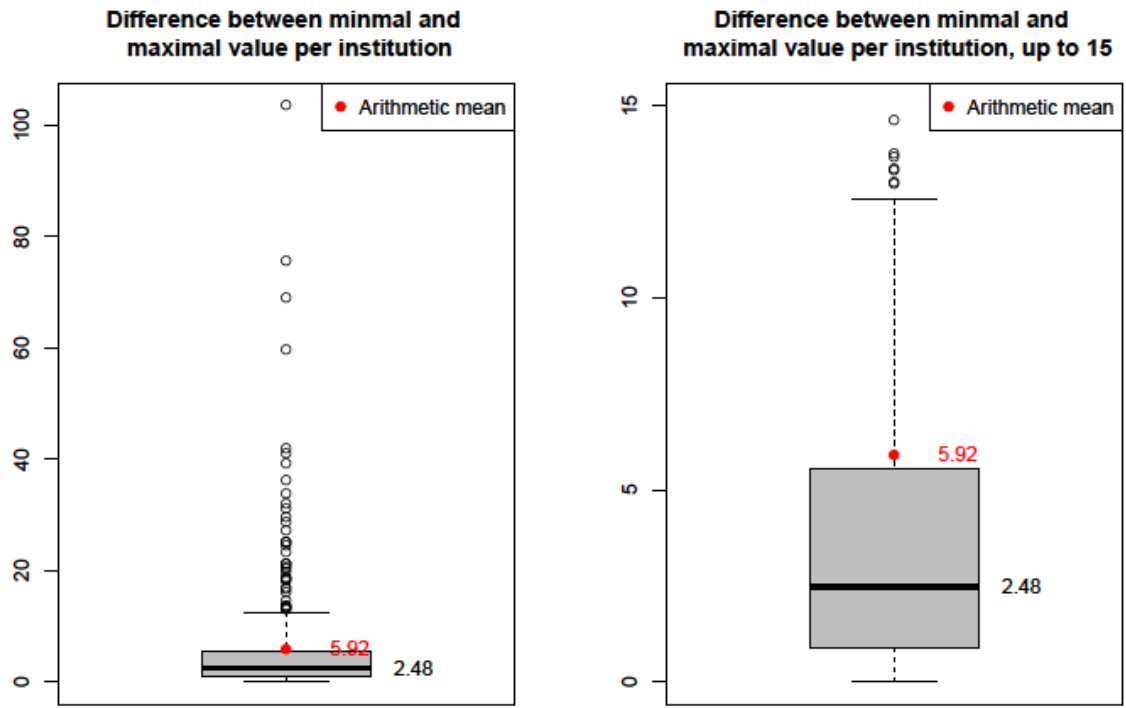

*Figure 13: Differences between minimal and maximal value per institution*

Table 8 provides an overview of the universities with the largest differences ($PU_{diff}$) between the cost sharing model with the lowest ($PU_{min}$) and the highest publication units ($PU_{max}$) to be paid by that institution.

| Institution | $PU_{min}$ | $PU_{max}$ | $PU_{diff}$ | Cost sharing models |
|---|---|---|---|---|
| Deutsches Krebsforschungszentrum (DKFZ) | 275.00 | 378.51 | 103.51 | 2 \| 3 |
| Ruprecht-Karls-Universität Heidelberg | 1,405.14 | 1,480.67 | 75.53 | 1 \| 3 |
| Humboldt-Universität zu Berlin | 291.73 | 360.85 | 69.12 | 2 \| 3 |
| Johann Wolfgang Goethe-Universität | 665.86 | 725.67 | 59.81 | 2 \| 3 |
| Universitätsklinikum Schleswig-Holstein | 114.67 | 156.51 | 41.84 | 2 \| 3 |
| Charité - Universitätsmedizin Berlin | 832.83 | 873.9 | 41.07 | 2 \| 3 |
| University of Freiburg | 851,76 | 891,08 | 39.32 | 1 \| 3 |
| University of Erlangen Nürnberg | 880,64 | 916.83 | 36.19 | 2 \| 3 |

*Table 8: Institutions with the largest differences between the cost sharing models with min. and max. publication units*

A closer look at the outliers reveals three results worth highlighting. First, and for the institutions with the largest differences ($PU_{diff}$), the APC that would have to be spent for $PU_{max}$ instead of $PU_{min}$ would sum up to a considerable amount of money. Taking the average APC for publications in Gold OA journals of German institutions as a proxy, Deutsches Krebsforschungszentrum would have to spend additional 159,405 € in the case of $PU_{max}$ instead of $PU_{min}$, Ruprecht-Karls-University Heidelberg additional 116,316€ and Humboldt University additional 106,445€ in the period 2014-2018. Second, the outliers with the largest differences ($PU_{diff}$) are without exception universities with a medical faculty or research institutes in the field of medical science. From the analysis presented here, no conclusion can be drawn whether this result is a coincidence or if it is caused by systematic characteristics of the composition of author teams in this field. Third, it is striking that for all 8 institutions the largest differences occur between a cost sharing model that applied full counting attribution of costs and a model that applies fractional counting.

This observation raises the question whether the largest differences between the cost sharing models only occur in the group of outliers or if it also holds for the other institutions. Therefore, the complete dataset including 361 institutions is analyzed with regard to which cost sharing models differ most, and which are most similar. Figure 14 to 16 present the results and confirm the analysis of the outliers. The two models with the largest differences are the equal shares (3) and the reprint author model (2) followed by the equal shares model (3) and the first author model (1). On the other side of the spectrum, the two fractional counting models 4a and 4b lead to the most similar attribution of costs.

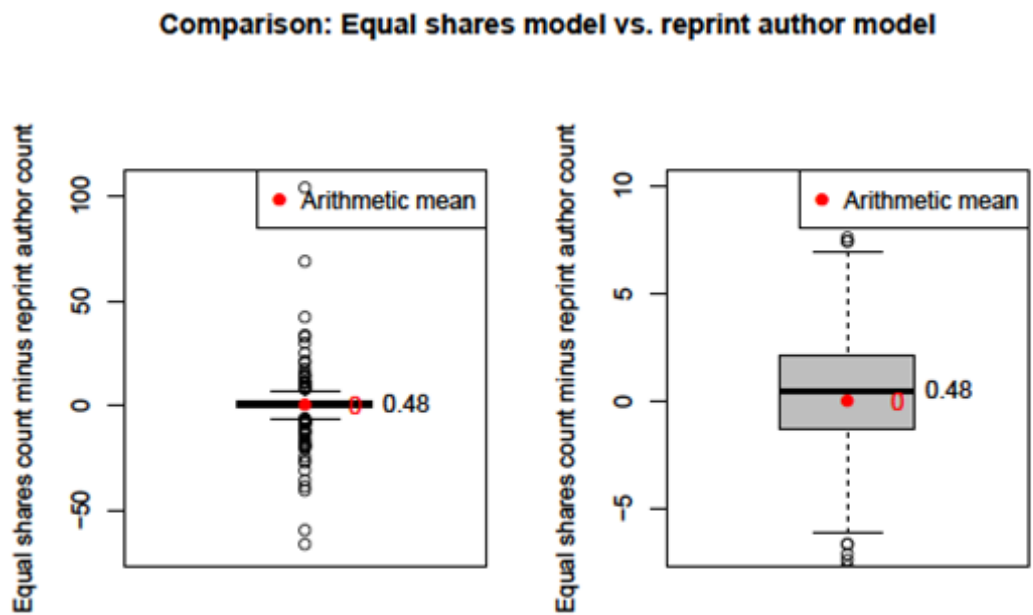

*Figure 14: Differences between equal shares model and reprint author count model*

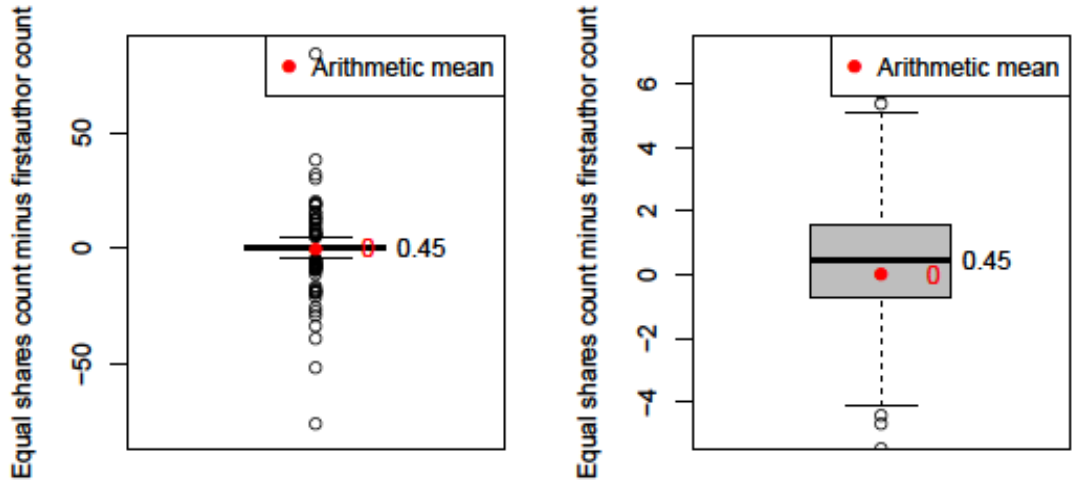

*Figure 15: Differences between equal shares count and first author count model*

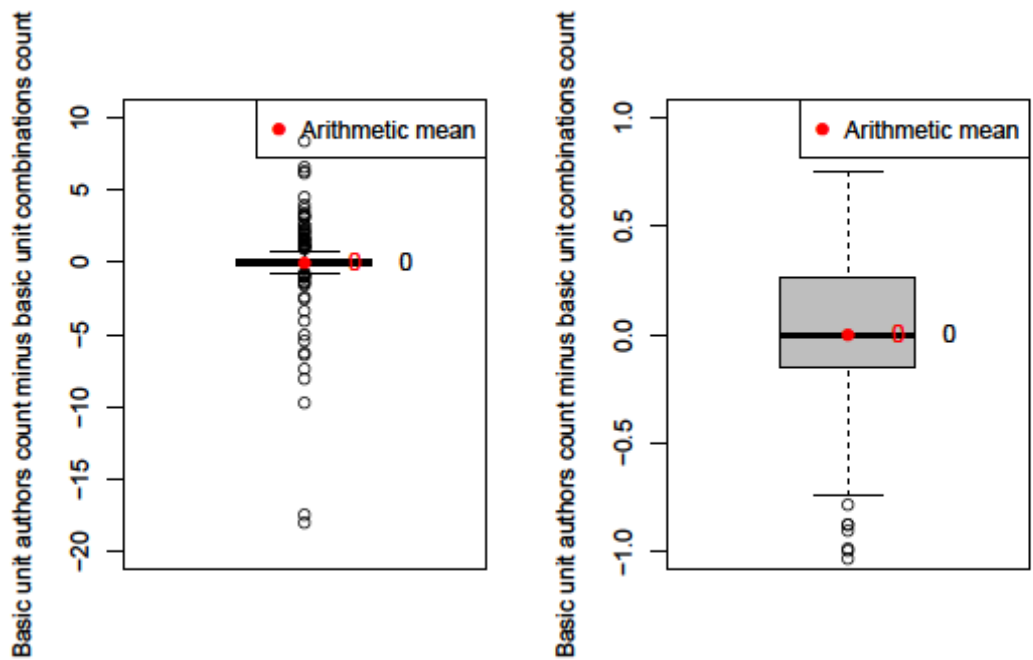

*Figure 16: Differences between author-institution-combination count and basic unit authors count model*

In a concluding step, the distributional effects are also calculated on the level of the five sectors of the German research system. Table 8 provides an overview of the costs for all cost sharing models for each sector. The differences between the models are also considerable at this level of analysis. In the case of universities, the difference between the model with the highest and the lowest costs sums up to 153.8 publication units, followed by Helmholtz with 100.6, WGL with 77.2, MPG with 62.3, and finally Fraunhofer with 44.1 publication units.

| Sector | first author | reprint author | equal share | basic unit: author-institution-combination | basic unit: author |
|---|---|---|---|---|---|
| Universities | 27,841.24 | 27,773.89 | 27,687.46 | 27,746.71 | 27,764.56 |
| MPG | 1,505.37 | 1,533.20 | 1,567.67 | 1,537.27 | 1,538.90 |
| HGF | 2,527,18 | 2,512.83 | 2,613.47 | 2,579.55 | 2,557.01 |
| WGL | 1,291.95 | 1,330.83 | 1,253.03 | 1,275.82 | 1,282.11 |
| Fraunhofer | 303.17 | 318.16 | 347.29 | 329.71 | 326.42 |

*Table 9: Differences between cost sharing models per sector*

## 7. Conclusion

The most important result of this analysis is that the type of model that is applied to attribute costs for article processing charges matters – not for all, but at least for some institutions. For the overwhelming majority the application of a cost sharing model is not an issue, as they roughly lead to the same amount that has to be spent for APC. In contrast, the institutions identified as outliers might have a strong preference for a particular model as the models will lead to a considerable amount of money that either has to be spent or that can be saved. The results of the analysis have to be put into the following context. A 5-year time span was chosen in order to come to reliable results and to analyze a significant number of OA publications per institution. Therefore, it has to be considered that – unless otherwise indicated – the results refer to the overall period and should not be confused with yearly indicators and costs. Nevertheless, two characteristics of the analysis underline the importance of the hitherto overlooked issue of different possible cost sharing models and possible financial effects of a Gold OA transformation: First, the analysis was restricted to German publications only. If international publications have the same structure of co-authorship as the set of publications considered here, the differences between the models are expected to be larger. Second, the analysis was restricted to publications in Gold OA journals. Assuming that such publications have the same structure of co-authorship as non-OA, it is to be expected that the differences between the models also increase in the course of the transformation towards Gold OA publishing.

Especially if money is limited, not only the question of the financial burdens of institutions with a strong publication record (Lara 2014; Solomon/Björk 2016: 2) may become a bigger issue, but also the model applied for the attribution of costs. In that case the analysis undertaken here points to a possible lines where a conflict is most likely to arise: Between models that attribute costs on the basis of full counting and those that are based on fractional counting.

# References

Abendiyarandi, Neda; Mayr, Philipp (2019). The State of Open Access in Germany: An Analysis of the Publication Output of German Universities. arxiv: 1905:00011v3.

Aksnes, Dag W.; Schneider, Jesper W.; Gunnarsson, Magnus (2012). Ranking national research systems by citation indicators. A comparative analysis using whole and fractional counting methods. *Journal of Informetrics*, Vol. 6, No. 1, pp. 36–42.

Appel, Andre Luiz; Albagli, Sarita (2019). The adoption of Article Processing Charges as a business model by Brazilian Open Access journals. *TransInformação Campinas*, Vol. 31:e180045.

Archambault, Éric; Amyot, Didier; Deschamps Philippe; Nicol, Aurore; Provencher Françoise; Rebout, Lise; Roberge, Guillaume (2014). *Proportion of open access papers published in peer-reviewed journals at the European and world levels–1996–2013. European Commission*. Available at: http://science-metrix.com/sites/default/files/science-metrix/publications/d_1.8_sm_ec_dg-rtd_proportion_oa_1996-2013_v11p.pdf.

Asai, Sumiko(2019). Determinants of Article Processing Charges for Medical Open Access Journals. *Journal of Electronic Publishing,* Vol. 22, No. 1. DOI: 10.3998/3336451.0022.103.

Björk, Bo-Christer; Solomon, David (2014). Developing an Effective Market for Open Access Article Processing Charges. Final Report to a consortium of research funders comprising Jisc, Research Libraries UK, Research. Councils UK, the Wellcome Trust, the Austrian Science Fund, the Luxembourg National Research Fund and the Max Planck Institute for Gravitational Physics. Available at: https://fwf.ac.at/fileadmin/files/Dokumente/Downloads/Dev_Effective_Market_OA_ Article_Processing_Charges.pdf.

Björk, Bo-Christer; Solomon, David (2015). Article processing charges in OA journals: relationship between price and quality. *Scientometrics* , Vol. 103, No. 2, pp. 373–385. DOI: 10.1007/s11192-015-1556-z.

Bruns, Andre; Lenke, Christopher; Schmidt, Constanze; Taubert, Niels (2019). *ISSN-Matching of Gold OA Journals (ISSN-GOLD-OA) 3.0*. Bielefeld University. DOI: 10.4119/unibi/2934907.

Cole, Jonathan R.; Cole, Stephen (1973). *Social Stratification in Science*. Chicago/London: The University of Chicago Press.

Crawford, Walt (2015). *The Gold OA Landscape 2011-2014*. Liversmore: Cites & Insights Books.

Crawford, Walt (2017). *GOAJ2. Gold Open Access Journals 2011-2016*. Liversmore: Cites & Insights Books.

Cuhls, Kerstin; Bunkowski, Alexander ; Behlau, Lothar (2012). Fraunhofer future markets: From global challenges to dedicated, technological, collaborative research projects, *Science and Public Policy*, Volume 39, No. pp. 232–244, https://doi.org/10.1093/scipol/scs018.

Finch, Janet (2012). Accessible, sustainability, excellence. How to expand access to research publications. Report of the Working Group on Expanding Access to Published Research Findings. Available at: http://www.researchinfonet.org/wp-content/uploads/2012/06/Finch-Group-report-FINAL-VERSION.pdf.

Fuchs, Christian; Sandoval, Marisol. (2013). The diamond model of open access publishing: Why policy makers, scholars, universities, libraries, labour unions and the publishing world need to take non-commercial, non-profit open access serious. *TripleC: Communication, Capitalism & Critique*, Vol. 11, No. 2, pp. 428–443.

Gargouri, Yves; Larivière, Vincent; Gingras, Yves; Carr, Leslie; Harnad, Steven (2012), Green and gold open access percentages and growth by disciplines. ArXiv preprint. arXiv: 1206.3664.

Heinze, Thomas; Kuhlmann, Stefan (2008): Across institutional boundaries? Research collaboration in German public sector nanoscience. *Research Policy, Vol.* 37, No. 5, pp. 888–899. DOI:10.1016/j.respol.2008.01.009.

Hohn, Hans-Willy (2010). Außeruniversitäre Forschungseinrichtungen. pp. 457–477. In: Simon, Dagmar; Knie, Andreas; Hornbostel, Stefan (Eds.) *Handbuch Wissenschaftspolitik*. Wiesbaden: VS Verlag für Sozialwissenschaften.

Hohn, Hans-Willy (2016). Governance-Strukturen und institutioneller Wandel des außeruniversitären Forschungssystems Deutschlands. pp. 549–532. In: Simon, Dagmar; Knie, Andreas; Hornbostel, Stefan; Zimmermann, Karin (Eds.) *Handbuch Wissenschaftspolitik*. 2$^{nd}$ . Edition. Wiesbaden: Springer VS.

Huang, Ding-Wei (2015). Temporal evolution of multi-authored papers in basic science from 1960 to 2010. *Scientometrics*, Vol. 105, No. 3, pp. 2137–2147.

Kupferschmidt, Kai (2019). Groundbreaking deal makes large number of Germany studies free to public. *Science*. DOI:10.1126/science.aaw6836.

Laakso, Mikael; Welling, Patrik; Bukvova, Helena; Nyman, Linus; Björk, Bo-Christer, Hedlund, Turid (2011): The development of Open Access Journal Publishing from 1993 to 2009. *PLoS ONE*, Vol. 6, No. 6, e20961.

Lara, Kate 2014: Open Access Library Survey. An investigation of the role of libraries in open access funding and support within institutions. Report. Publisher Communication Group. Available at: http://www.pcgplus.com/wp-content/uploads/2014/09/PCG-Open-Access-Library-Survey-2014.pdf.

Martín-Martín, Alberto; Costas, Rodrigo; van Leeuwen, Thed, Lópeu-Cózar, Emilio Delgado (2018). Evidence of open access of scientific publications in Google Scholar. A large-Scale analysis. *Journal of Informetrics*, Vol. 12, No.3, pp. 819–841.

Mellon Foundation (2016): Pay it forward. Investigating a Sustainable Model of Open Access Article Processing Charges for Large North American Research institutions. Available at: https://www.library.ucdavis.edu/wp-content/uploads/2018/11/ICIS-UC-Pay-It-Forward-Final-Report.rev_.7.18.16.pdf.

Morrison, Heather; Salhab, Jihane; Calvé-Genest, Alexis; Horava, Tony (2015). Open Access Article Processing Charges: DOAJ Survey May 2014. *Publications*, Vol. 3, No. 1, pp. 1–16, DOI: 10.3390/publications3010001.

Perianes-Rodriguez, Antonio; Waltman, Ludo; Van Eck, Nees Jan (2016): Constructing bibliometric networks: A comparison between full and fractional counting. *Journal of Informetrics*, Vol. 10, No. 4, pp. 1178–1195. DOI: 10.1016/j.joi.2016.10.006.

Pieper, Dirk; Broschinski, Christoph (2018), OpenAPC: a contribution to a transparent and reproducible monitoring of fee-based open access publishing across institutions and nations. *Insights*, Vol. 31, 39, pp. 1–18. DOI: 10.1629/uksg.439.

Pinfield, Stephen; Salter, Jennifer; Bath Peter A. (2015). The "Total Costs of Publication" in a Hybrid Open-Access Environment: Institutional Approaches to Funding Journal Article-Processing Charges in Combination with Subscriptions. *Journal of the Association for Information Sciences and Technology*, Vol. 67, No. 7, pp. 1751–1766. DOI: 10.1002/asi.23446.

Piwowar, Heather; Priem, Jason; Larivière, Vincent; Alperin, Juan Pablo; Mattias, Lisa; Norlander, Bree; Farley, Ashley; West, Jevin; Haustein, Stefanie (2018), The state of OA: a large-scale analysis of the prevalence and impact of Open Access articles. *PeerJ* 6:4375, DOI: 10.7717/peerj.4375.

Prosser, David C. 2003. From here to there: a proposed mechanism for transforming journals from closed to open access. *Learned Publishing*, Vol. 16, No. 3, pp. 163–166. DOI: 10.1087/095315103322110923.

Rimmert, Christine; Schwechheimer, Holger; Winterhager, Matthias (2017). *Disambiguation of author addresses in bibliometric databases – technical report.* Bielefeld: Bielefeld University, Institute for Interdisciplinary Studies of Science (I²SoS); 2017.

Schimmer, Ralf; Geschuhn, Kai Karin; Vogler, Andreas (2015). Disrupting the subscription journals' business model for the necessary large-scale transformation to open access. A Max Plack Digital Library Open Access Policy White Paper. DOI: 10.17617/1.3

Solomon, David; Björk, Bo-Christer (2012). Pricing principles used by scholarly open access publishers. *Learned Publishing*, Vol. 25, No. 2, pp. 132–137. DOI: 10.1087/20120207.

Solomon, David; Björk, Bo-Christer (2016). Article Processing charges for open access publications – the situation for research intensive universities in the USA and Canada. *PeerJ* 4:e2264, DOI: 10.7717/peerj.2264.

Taubert, Niels (2019). *Open-Access-Transformation. Abschätzung der zur Verfügung stehenden Mittel für Publikationsgebühren in Forschungsorganisationen – Verfahren, Ergebnisse und Diskussion. Bielefeld University: Forschungsbericht*. DOI: 10.4119/unibi/2933620.

University of California Libraries (2016). Pay it Forward. Investigating a Sustainable Model of Open Access Article Processing Charges for Large North American Research Institutions. Available at: https://www.library.ucdavis.edu/wp-content/uploads/2018/11/ICIS-UC-Pay-It-Forward-Final-Report.rev_.7.18.16.pdf.

Van Hooydonk, Guido (1997). Fractional Counting of Multiauthored Publications: Consequences for the Impact of Authors. *Journal of the American Society for Information Science*, Vol. 48, No. 10, pp. 944–945.

Wilson, T.D. (2007, April 19). Re: Bundesrat decision [Msg. 1078]. Message posted to http://threader.ecs.soton.ac.uk/lists/boaiforum.

Wohlgemuth, Michael; Rimmert, Christine; Taubert, Niels (2017). *Publikationen in Gold-Open-Access-Journalen auf globaler und europäischer Ebene sowie in Forschungsorganisationen*. Bielefeld: Universität Bielefeld. DOI: 10.13140/RG.2.2.33235.89120.

Wohlgemuth, Michael; Rimmert, Christine; Winterhager, Matthias (2016). *ISSN-Matching of Gold OA Journals (ISSN-GOLD-OA)*. Bielefeld University. DOI:10.4119/unibi/2906347.